\pdfoutput=1
\documentclass[PAPER, atlasdraft=false, cernpreprint, USenglish, texmf, orcidlogo]{atlasdoc}
\usepackage{atlaspackage}
\usepackage{animate}
\usepackage{atlasbiblatex}

\usepackage{atlasphysics}
\graphicspath{{logos/}{figures/}}

\usepackage{EXOT-2016-20-PAPER-defs}

\AtlasTitle{Search for low-mass dijet resonances using trigger-level jets with the ATLAS detector in $pp$ collisions at $\sqrt{s}=13$~\TeV}

\AtlasAbstract{%
Searches for dijet resonances with sub-\TeV\ masses using the ATLAS detector at the Large Hadron Collider can be statistically limited by the bandwidth available to inclusive single-jet triggers, whose data-collection rates at low transverse momentum are much lower than the rate from Standard Model multijet production. This Letter describes a new search for dijet resonances where this limitation is overcome by recording only the event information calculated by the jet trigger algorithms, thereby allowing much higher event rates with reduced storage needs. The search targets low-mass dijet resonances in the range 450--1800~\GeV. The analyzed dataset has an integrated luminosity of up to \integLumi and was recorded at a center-of-mass energy of 13~\TeV. No excesses are found; limits are set on Gaussian-shaped contributions to the dijet mass distribution from new particles and on a model of dark-matter particles with axial-vector couplings to quarks.}

\author{The ATLAS Collaboration}

\AtlasRefCode{EXOT-2016-20}

\PreprintIdNumber{CERN-EP-2018-033}

\AtlasJournalRef{Phys.\ Rev.\ Lett. 121 (2018) 081801}
\AtlasDOI{10.1103/PhysRevLett.121.081801}

\hypersetup{pdftitle={ATLAS document},pdfauthor={The ATLAS Collaboration}}

\begin{document}

\maketitle

\section{Introduction}

If new particles beyond those of the Standard Model (SM) are directly produced in proton--proton ($pp$) collisions at the Large Hadron Collider (LHC), they must interact with the constituent partons of the proton, and can therefore also decay into the same partons, resulting in two-jet final states. Quantum chromodynamics (QCD) predicts that dijet events have an invariant mass distribution (\mjj{}) that falls smoothly, whereas a new state decaying to two partons would emerge as a localized excess in the distribution.

\begin{figure}[bht]
\centering
\includegraphics[width=0.45\linewidth]{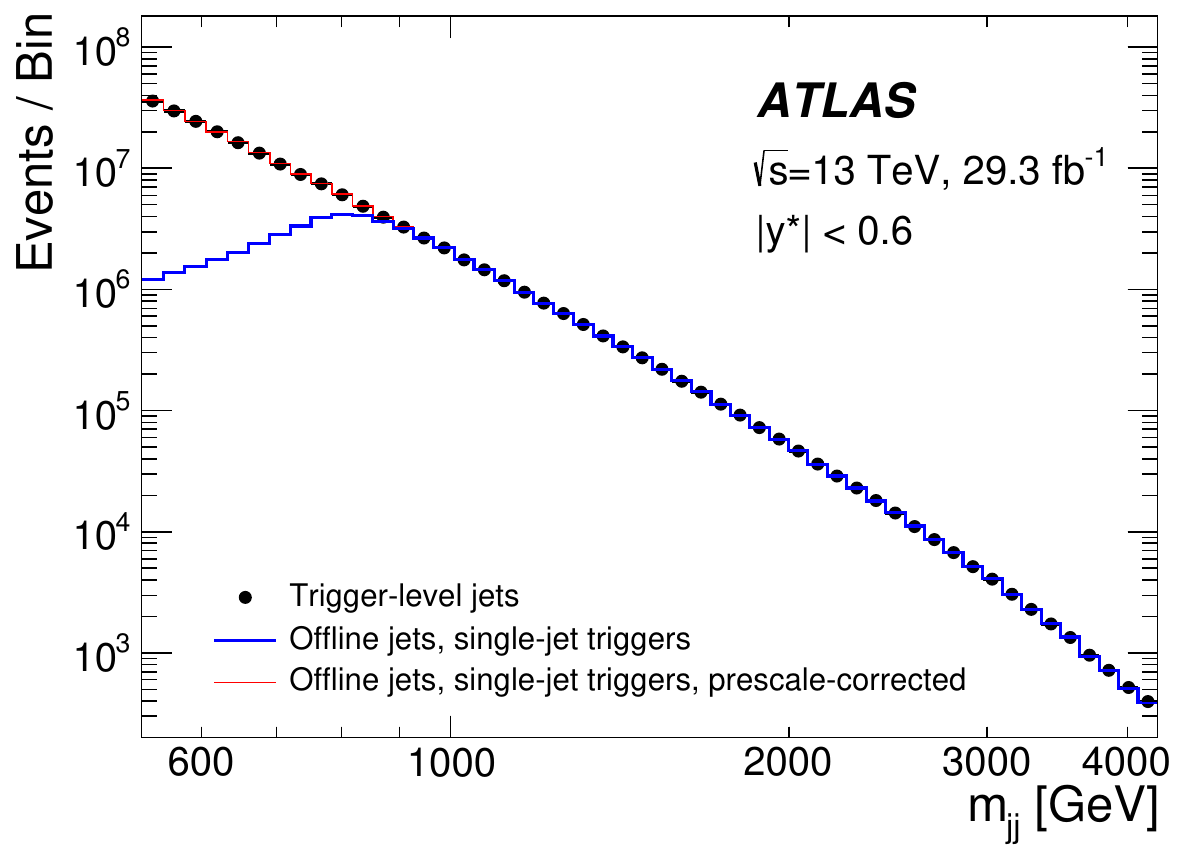}
\caption{Comparison between the number of dijet events in the data used by this analysis (black points), the number of events selected by any single-jet trigger (thicker, blue line), and the events selected by single-jet triggers but corrected for the trigger prescale factors (thinner, red line) as a function of the dijet invariant mass (\mjj). The definition of \ystar\ is $(y_1-y_2)/2$, where $y_1$ and $y_2$ are the rapidities of the highest- and second-highest-\pT\ jets.}
\label{fig:mjj-triggerPrescale}
\end{figure}

Traditional dijet searches at the LHC focus on the production of heavy particles with masses above 900~\GeV~\cite{Harris:2011bh,CMS-EXO-15-001,EXOT-2016-21}.
LHC searches for lighter resonances with small production cross-sections have been hampered by restrictions in the data-taking rate of the ATLAS and CMS detectors. Single-jet triggers with a jet \pT{} threshold below roughly 380~\GeV\ are prescaled, a procedure whereby only a fraction of the events passing the trigger are recorded, hence dijet events with an invariant mass below 1~\TeV\ are largely discarded by the trigger system, as indicated in Figure~\ref{fig:mjj-triggerPrescale}. Therefore, despite the large number of $pp$ collisions produced by the LHC, traditional ATLAS and CMS searches are less sensitive to dijet resonances below 900~\GeV\ than searches at the SPS and Tevatron colliders~\cite{Arnison:1983dk,Albajar:1988rs,Bagnaia1984283, CDF:2009DijetSearch,Dobrescu:2013cmh,Chala:2015ama}. Alternative trigger strategies to search for low-mass resonances include selecting events with jets recoiling against either an energetic photon or an additional energetic jet~\cite{ATLAS-CONF-2016-070,CMS-EXO-17-001,EXOT-2017-01}, or selecting events with decays to heavy-flavor jets~\cite{EXOT-2016-33,CMS-EXO-16-057}. In these cases, additional features in the events reduce the data-taking rates, reducing the sensitivity to low-mass resonances.

This Letter describes an innovative data-taking approach to access the invariant mass region below 1~\TeV; only a reduced set of information from the trigger system is recorded and subsequently analyzed. The trigger-object-level analysis (TLA) approach allows jet events to be recorded at a peak rate of up to twice the total rate of events using the standard approach, while using less than 1\% of the total trigger bandwidth~\cite{TRIG-2016-01}.
This strategy is used in dijet resonance searches in $\sqrt{s}=$8 and 13~\TeV\ LHC $pp$ collision data by the CMS Collaboration~\cite{CMS-EXO-14-005,Sirunyan:2018xlo}, and it is used by the LHCb Collaboration (e.g. \cite{Aaij:2015rla}).
The analysis presented here uses \integLumi of $\sqrt{s}=13$~\TeV\ $pp$ collision data recorded in 2016 by the ATLAS detector.

\newcommand{\AtlasCoordFootnote}{%
ATLAS uses a right-handed coordinate system with its origin at the nominal interaction point (IP) in the center of the detector and the $z$-axis along the beam pipe. The $x$-axis points from the IP to the center of the LHC ring, and the $y$-axis points upwards. Cylindrical coordinates $(r,\phi)$ are used in the transverse plane, $\phi$ being the azimuthal angle around the $z$-axis. The pseudorapidity is defined in terms of the polar angle $\theta$ as $\eta = -\ln \tan(\theta/2)$. The rapidity, $y$, is defined as $\frac{1}{2}\ln[(E+p_z)/(E-p_z)]$, where $E$ denotes the energy of the jet and $p_z$ the momentum component of the jet along the beam direction. Angular distance is measured in units of $\Delta R \equiv \sqrt{(\Delta\eta)^{2} + (\Delta\phi)^{2}}$.
}

\section{ATLAS detector and data sample}

The ATLAS detector~\cite{PERF-2007-01} is a multipurpose detector with a forward-backward symmetric cylindrical geometry and nearly $4\pi$ coverage in solid angle,\footnote{\AtlasCoordFootnote} consisting of tracking detectors, calorimeters and a muon spectrometer. In the pseudorapidity region $|\eta| < 3.2$, high-granularity lead and liquid-argon (LAr) electromagnetic sampling calorimeters are used. A steel and scintillator hadronic tile calorimeter provides coverage in the range $|\eta| < 1.7$. Hadronic calorimetry in the endcap region, $1.5<|\eta|<3.2$, and electromagnetic and hadronic calorimetry in the forward region, $3.1<|\eta|<4.9$, are provided by LAr sampling calorimeters. A two-level trigger system is used to select events for offline storage~\cite{TRIG-2016-01}. A first-level (L1) trigger based on dedicated hardware identifies jets from $\Delta \eta \times \Delta \phi = 0.2 \times 0.2$ calorimeter segments with a sliding-window algorithm. Events passing the L1 trigger are processed by a software-based high-level trigger (HLT). The HLT system reconstructs jets using the anti-$k_t$ algorithm~\cite{Cacciari:2008gp,Cacciari:2006} with radius parameter $R = 0.4$. The inputs to this algorithm are groups of contiguous calorimeter cells (topological clusters), in which each cell's inclusion is based on the significance of its energy deposit over calorimeter noise~\cite{PERF-2014-07}. Jet four-momenta are computed by summing over the four-momenta of the topological clusters that compose the jet, with each cluster pointing to the center of the ATLAS detector and being treated as massless. The HLT jet reconstruction uses the same techniques that the ATLAS offline jet reconstruction applies to similar inputs from recorded data events that include the full detector information~\cite{TRIG-2016-01}.

After execution of the HLT jet algorithm, only \triggerLevel jets with $\pt{}>20$~\GeV\ are stored. The stored information includes the four-momentum of each jet and a set of calorimeter variables characterizing the jet~\cite{ATLAS-CONF-2015-029}, such as information about the jet quality and structure. The average size of these events is less than 0.5\% of the size of full offline events that contain all detector information. For this analysis, all events containing at least one L1 jet with $\ET>100$~\GeV\ are selected and recorded, corresponding to a total luminosity of \integLumi{}. Events containing a L1 jet with $\ET>75$~\GeV\ are also selected, in a subset of this data corresponding to \integLumiSmall{} that was collected at the start of the data-taking period. Events with at least one L1 jet with $\ET>100$~\GeV\ are therefore included in both datasets.

\section{Calibration procedure}

After the events are recorded, the \triggerLevel jet energy and direction are corrected to those of simulated particle-level jets built from stable particles with a lifetime longer than 10 ps, excluding muons and neutrinos. Before any calibration, the jet \pt\ response, defined as the \pt\ ratio of a \triggerLevel jet to the same jet\footnote{The \triggerLevel and offline jets are matched within a radius of $\Delta R = 0.4$.} reconstructed offline (offline jet), is between 0.95 and 1.05 in the \pt\ range considered by this analysis. Since the energy scale for \triggerLevel and offline jet are very similar, the \triggerLevel jet calibration employs the same procedure and constants as derived for offline jets~\cite{PERF-2016-04}, with some modifications to account for the unavailability of tracking information for \triggerLevel jets.

\begin{figure}[h!]
\centering
\includegraphics[width=0.80\columnwidth]{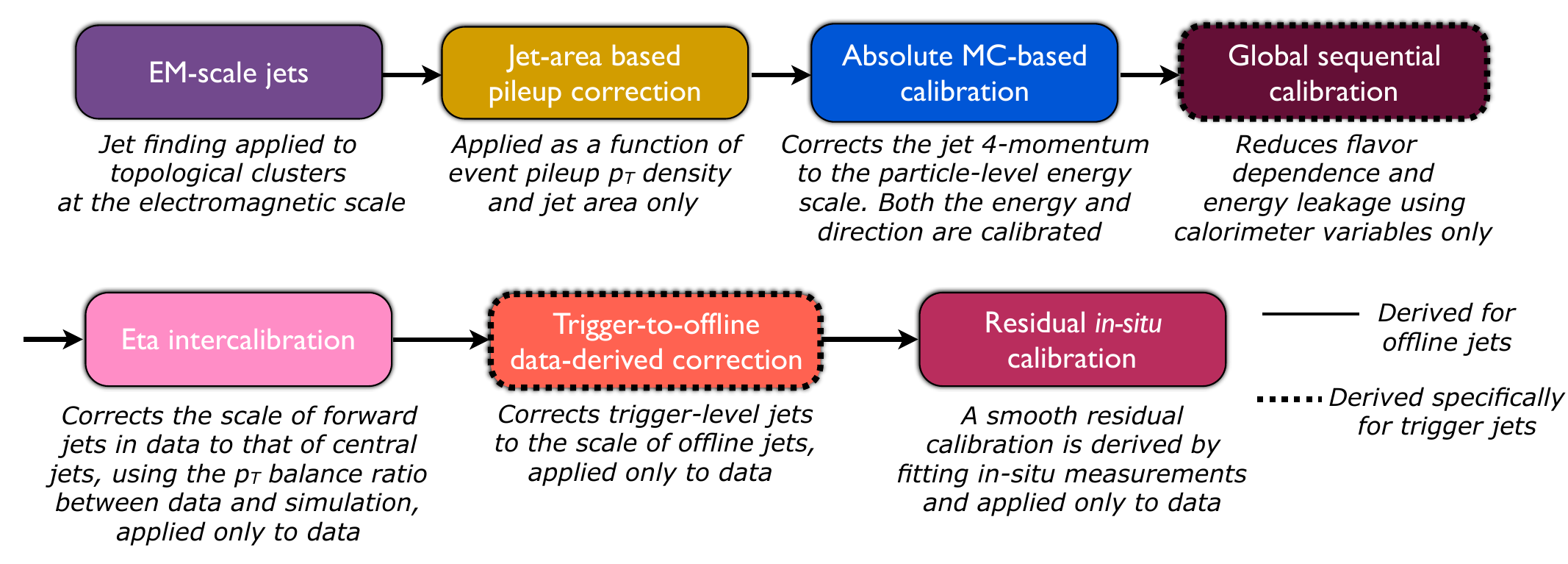}
\caption{Calibration stages for EM-scale \triggerLevel jets, each applied to the four-momentum of the jet. MC refers to the simulation.}
\label{fig:CalibrationScheme}
\end{figure}

In the calibration procedure, summarized in Figure~\ref{fig:CalibrationScheme}, an event-by-event jet-area-based calibration~\cite{PERF-2014-03} is used to correct for contributions from additional proton--proton interactions (pileup) in the same and neighboring crossings of proton bunches. Then, the simulation-based calibration derived for offline jets is applied to \triggerLevel jets to correct both jet energy and direction.
Next, calorimeter-based variables are used to reduce the dependence on the \triggerLevel jet flavor and to minimize the impact of energy leakage. Only variables related to the \triggerLevel jet energy fractions in the electromagnetic and hadronic calorimeters and the minimum number of calorimeter cells containing 90\% of the \triggerLevel jet energy are used here since track-based variables, which are normally used in the offline calibration, are not available. With this correction, the \triggerLevel jet energy resolution is improved by 8\% at jet \pt\ values of 85~\GeV\ and up to 40\% for jet \pt\ values of 1~\TeV\ relative to the previous calibration step. Next, the calibration corrections that restore the relative calibration between central and forward jets in data and simulation are derived for offline jets and applied to \triggerLevel jets. After these calibration steps, any residual difference between \triggerLevel jets and offline jets is accounted for in a dedicated correction, based on the \pt\ response and derived from data in bins of jet $\eta$ and \pT{}. The size of this correction is on average 1\%, with values reaching up to 4\% in the endcap regions of the calorimeter.

\begin{figure}
\centering
\includegraphics[width=0.5\linewidth]{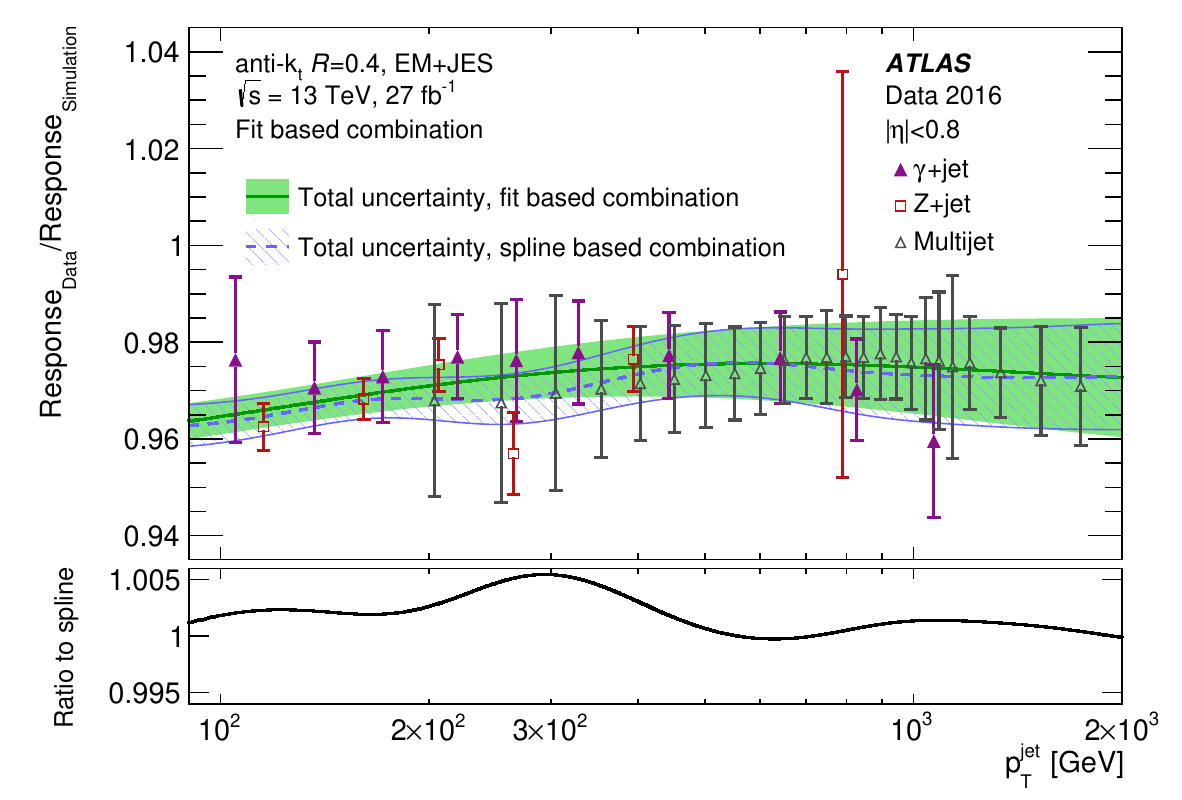}
\caption{The in situ calibration in the range 85~\GeV $<$ jet \pT\ $< $ 2~\TeV, for both the spline (dashed line) and fitted (solid line) combination methods, as described in the text. Data points from the three input measurements are overlaid. The lower panel shows the ratio of the two calibration curves.}
\label{fig:calib}
\end{figure}

Finally, an in situ calibration is obtained from the data-to-simulation ratio of the \pt\ balance between offline jets and well-calibrated objects against which the jets recoil. Three different types of well-calibrated objects are used to span the full \pT{} range of the jets: $Z$ bosons decaying to electrons or muons, photons, and multijets. A polynomial in $\log(\pT)$ is simultaneously fit to the three input measurements to combine them. The resulting curve is taken as the calibration correction to be applied to \triggerLevel jets. In deriving the final calibration curve the fit is chosen over the simple spline-based combination procedure used for offline jets in Ref.~\cite{PERF-2016-04}; this procedure is more robust against localized fluctuations in the jet \pT{} distribution that result in deviations from the expected smoothly falling invariant mass spectrum. Any dependence of the final mass spectrum on the parameterization of the jet energy scale calibration is tested by comparing different parameterizations on the data as well as on simulations. The fitted in situ calibration curve is compared to the spline-based smoothing procedure in Figure~\ref{fig:calib}. After the full calibration procedure, the energy of \triggerlevel jets is equivalent to that of offline jets to better than 0.05\% for invariant masses of 400~\GeV\ and their difference is negligible for invariant masses of 1~\TeV.

Energy scale and resolution uncertainties derived for offline jets~\cite{PERF-2016-04} are applied to \triggerlevel jets in the signal simulation, with additional uncertainties equivalent to the size of the final trigger-to-offline correction (1--3\%). The uncertainty due to the modeling of pileup effects and due to the jet parton flavor are derived specifically for \triggerlevel jets and are comparable to those of offline jets. The jet energy scale uncertainty for \triggerlevel jets at 200 GeV ranges from 1\% at $|\eta| < 0.8$ to 2\% in the region between the central and endcap regions ($1.0 < |\eta| < 1.5$).

\section{Event selection}

The dijet event selection for this analysis is similar to the one used in Ref.~\cite{EXOT-2016-21}. Events must contain at least two \triggerlevel jets with $|\eta|<2.8$. The leading \triggerlevel jet must have either $\pT > 185\ \GeV$ or $\pT > 220\ \GeV$ for the $\ET>75$~\GeV\ and $\ET>100$~\GeV\ L1 trigger selections, respectively; this ensures that the L1 triggers are fully efficient. The second leading jet must have a $\pT > 85\ \GeV$. Events that contain jets induced by calorimeter noise bursts, beam-induced background or cosmic rays are rejected using the same criteria as in Ref.~\cite{ATLAS-CONF-2015-029}, but omitting the track-based charged fraction selection, which is not available for \triggerlevel jets. This has a negligible effect for this analysis, since most of these backgrounds are already removed by requiring two central jets, as described below. The efficiency and purity of jets passing the selection are measured with a tag-and-probe method using data events with the full detector information. The \triggerlevel jet reconstruction efficiency is 100\% for jets with $\pT >85$~\GeV. The fraction of \triggerlevel jets that are not reconstructed and selected offline is below 0.1\%.

This analysis searches for a dijet resonance with a mass between 450~\GeV\ and 1800~\GeV{}. Two different selection criteria are used for different but overlapping ranges of the \mjj\ spectrum. To search for resonances with 700~\GeV~$ < \mjj <$~1800~\GeV, events are required to have $|\yStar| < 0.6$, where $\yStar = (y_1-y_2)/2$ and $y_1$ and $y_2$ are the rapidities of the highest- and second-highest-\pT\ \triggerlevel jets. To search for lower-mass resonances, with $\mjj > 450$~\GeV, events with $|\yStar| < 0.3$ are selected from the smaller data sample requiring a L1 jet with $\ET>75$~\GeV{}. The more stringent choice of $|\yStar| < 0.3$ selects higher-\pT{} jets at a given invariant mass and thus provides a mass distribution that is unbiased by the leading-jet selection from $\mjj = 450$~\GeV.

\section{Background estimation}

The invariant mass spectrum expected from SM dijet production is predicted to be smooth and falling. Prior dijet searches at various collision energies~\cite{Alitti:1990kw, CDF:2009DijetSearch,CMS-EXO-11-015,EXOT-2010-07,EXOT-2013-11,CMS-EXO-12-016} have found a variety of simple functional forms to describe this shape; however, given the statistical precision of the data and the wide invariant mass range covered by this search, none of the single, simple functional forms can provide a good description of the data.

The SM background distribution is determined using a sliding-window fit~\cite{EXOT-2016-21}, where a fitted functional form is evaluated at the bin at the center of a window, which then slides in one-bin steps along the \mjj\ distribution. The evaluated background estimates evaluated in each bin are then collated to form the final background estimate. The signal selection with $|\yStar| < 0.6$ uses a window size of 19 bins in the \mjj\ spectrum from 531~\GeV\ to 2080~\GeV, which spans 34 bins in total. The signal selection with $|\yStar| < 0.3$ uses a window size of 27 bins over a total of 40 bins, in the range 400 $ <\mjj\ <$ 2080~\GeV.
The bin sizes have been chosen according to the simulated invariant mass resolution: $\sigma_{\pT}/\pT=10.6/\pT \oplus 0.27/\sqrt{\pT} \oplus 0.039$.
The sliding window, however, can not be extended beyond the lower edge of the \mjj\ range used in each signal selection. Therefore, for the first 9 (13) bins in the $|\yStar| < 0.6$ ($|\yStar| < 0.3$) signal selection, which corresponds to one half of the window size, the window is fixed to the lower edge of the spectrum and instead of evaluating the fitted functional form at the window center, it is evaluated for each bin in turn. For invariant masses higher than the \mjj\ range used for the search, the window is allowed to extend beyond the range, to 2970 (3490)~\GeV\ for the $|\yStar| < 0.6$ (0.3) signal selection, %
and the fit is evaluated at the center of the window.

In each sliding window, three functional forms are fit to the data: a five-parameter function of the form
\begin{equation}
f(x) = p_1 (1 - x)^{p_2} x^{p_3 + p_4\ln x + p_5 (\ln x)^2} ,
\label{eq:backgroundfit}
\end{equation}
where $p_{i}$ are free parameters and $x \equiv \mjj/\sqrt{s}$; a four-parameter function,
which is the same as Eq.~(\ref{eq:backgroundfit}) but with $p_5 = 0$; and a four-parameter function used by the UA2 Collaboration~\cite{Alitti:1990kw}, defined as
\begin{equation}
f(x) = \frac{p_1}{x^{p_2}}{\mathrm{e}}^{-  p_3  x - p_4  x^2}.
\label{eq:UA2}
\end{equation}

The function used for each signal selection is the one that yields the best $\chi^2$ over the full fitted \mjj\ range. An alternative function is chosen to evaluate a systematic uncertainty. For the signal selection with $|\yStar| < 0.6$, Eq.~(\ref{eq:backgroundfit}) is used, yielding a $\chi^2$ $p-$value of 0.13, while the alternative function is the four-parameter function with a $\chi^2$ $p-$value of 0.11. For the signal selection with $|\yStar| < 0.3$, the four-parameter version of Eq.~(\ref{eq:backgroundfit}) yields the best $\chi^2$ $p-$value of 0.42 and the alternative function is Eq.~(\ref{eq:UA2}), with a $\chi^2$ $p-$value of 0.35.

The size of the sliding window is optimized to yield the best $\chi^2$ value for the full \mjj\ range while still being larger than the width of the expected signals and therefore insensitive to potential signal contributions. This latter requirement is checked by including signal models in pseudo-data samples and studying the dependence of the signal
sensitivity on different window sizes.

Systematic uncertainties in the estimate of the background used in setting limits include the uncertainty due to the choice of functional form and uncertainties in the fit parameter values. The effect of the choice of functional form is evaluated by comparing the nominal function to the alternative. The uncertainties in the fit parameter values are evaluated using pseudo-experiments, where the pseudo-data are drawn from Poisson fluctuations around the nominal background model.

\section{Results and limits}

\begin{figure}
\centering
\includegraphics[width=0.5\linewidth]{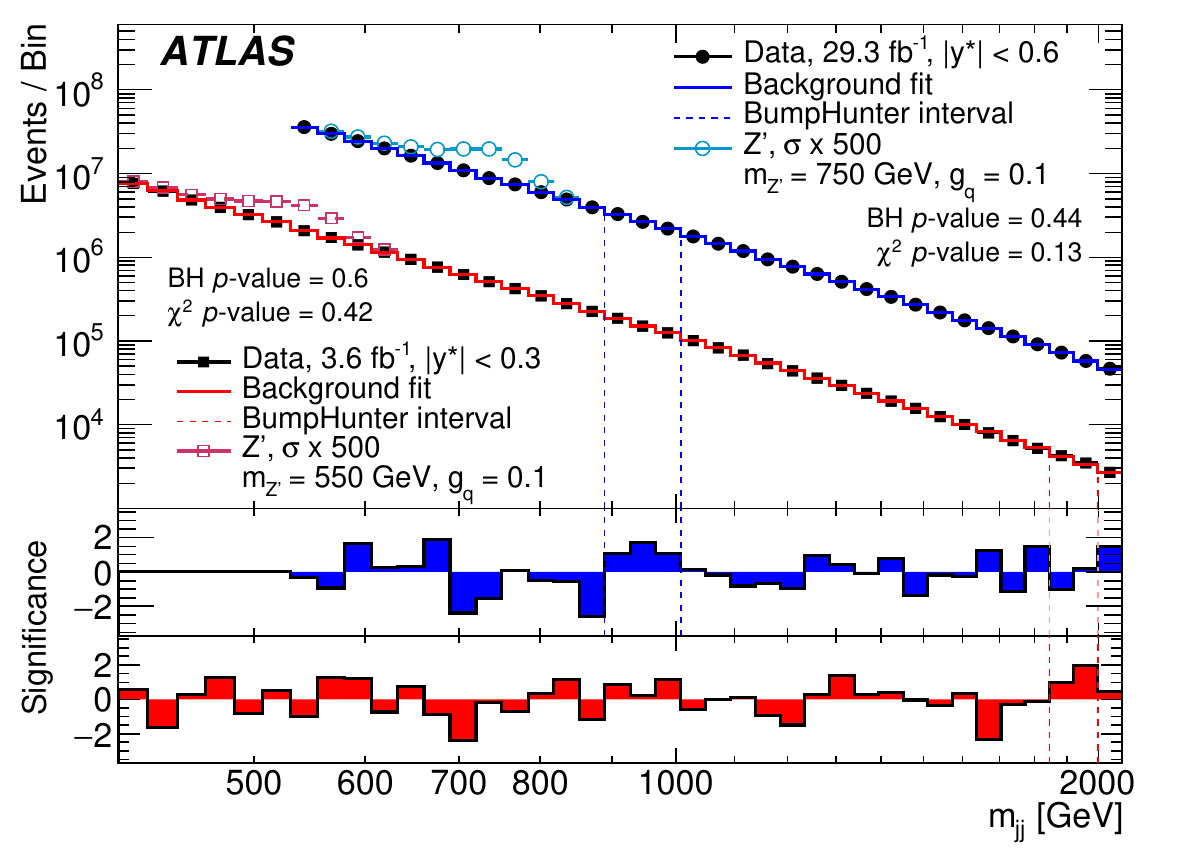}
\caption{The reconstructed dijet mass distribution (filled points) for events in the $|\yStar| < 0.3$ and $|\yStar| < 0.6$ signal regions. The solid lines depict the background estimate obtained by a sliding-window fit. Overall agreement between the background estimate and the data is quantified by the $\chi^2$ \textit{p}-value. The most discrepant localized excess in either signal region identified by the BumpHunter algorithm is indicated by the vertical lines. The open points show two possible signal models. The lower panels show the bin-by-bin significances of differences between the data and the background estimate, considering only statistical uncertainties (see Ref.~\cite{Choudalakis2012}).}
\label{fig:FullUnblinded}
\end{figure}

Figure~\ref{fig:FullUnblinded} shows the invariant mass distributions for dijet events in each signal region including the results from the sliding-window background estimates.
The $\chi^2$ \textit{p}-value of the overall background is 0.13 for the $|\yStar| < 0.6$ signal selection and 0.42 for the $|\yStar| < 0.3$ signal selection, indicating the data agrees well with the background estimate. The most discrepant interval identified by the \BumpHunter{} algorithm~\cite{Aaltonen:2008vt, Choudalakis:2011bh} is 889--1007~\GeV\ for events with $|\yStar| < 0.6$. Accounting for statistical uncertainties only, the probability of observing a deviation at least as significant as that observed in data, anywhere in this distribution, is 0.44 and corresponds to significance of 0.16 $\sigma$. Thus, there is no evidence of any localized excess.

Limits are set on both a leptophobic \Zprime{} simplified dark-matter model~\cite{Abercrombie:2015wmb} and a generic Gaussian model. The \Zprime\ simplified model assumes axial-vector couplings to SM quarks and to a Dirac fermion dark-matter candidate. No interference with the SM is simulated. Signal samples were generated so that the decay rate of the \Zprime{} into dark-matter particles is negligible and the dijet production rate and resonance width depend only on the coupling of the \Zprime{} to quarks, $g_q$, and the mass of the resonance, $m_{\Zprime}$~\cite{Chala:2015ama}. The model's matrix elements were calculated in \MADGRAPH~5~\cite{Alwall:2014hca} and parton showering was performed in \PYTHIA~8~\cite{pythia} with the A14 set of tuned parameters for underlying event~\cite{ATL-PHYS-PUB-2014-021} and NNPDF2.3 parton distribution functions~\cite{Ball:2012cx}.

The width of a \Zprime\ resonance with $g_q=0.10$, including parton shower and detector resolution effects, is approximately 7\%. Limits are set on the cross-section, $\sigma$, times acceptance, $A$, times branching ratio, $B$, of the model, and then displayed in the $(g_q, m_{\Zprime})$ plane.\footnote{Limits on the coupling are obtained accounting for the scaling of the signal cross-section with $g_q^2$.} The acceptance for a mass of 550~\GeV\ is 20\% for a \Zprime\ simplified model with $g_q=0.10$ for the $|\yStar| < $ 0.3 signal selection, and 41\% for a signal of mass equal to 750~\GeV\ for the $|\yStar| < $ 0.6 signal selection.

Limits are also set on a generic model where the signal is modeled as a Gaussian contribution to the observed \mjj\ distribution. For a given mean mass, $m_G$, four different Gaussian widths are considered: a width equal to the simulated mass resolution (which ranges between 4\% and 6\%), and the fixed fractions 5\%, 7\% and 10\% of $m_G$. As the width increases, the expected signal contribution is distributed across more bins. Wider signals are therefore less affected by statistical fluctuations from the data in a single bin. The results can be used to set limits on models of new phenomena besides that of the \Zprime{} simplified model and are applicable when the resonance is sufficiently narrow and the parton distribution function and non-perturbative effects can be safely truncated or neglected, as described in Ref.~\cite{EXOT-2013-11}. These criteria are often met if the \mjj\ distribution for a signal approaches a Gaussian distribution after applying the kinematic selection criteria of the resonance analysis, so that 95\% of the signal lies within 20\% of the Gaussian mean mass. Models of new resonances with an intrinsic width much smaller than 5\% of its mass should be compared to the results with a width equal to the experimental resolution. For models with a larger width, the limit that best matches their width should be used. More-detailed instructions can be found in Appendix A of Ref.~\cite{EXOT-2013-11}.

A Bayesian method is applied to the data and simulation of the signal models at a series of discrete masses to set 95\% credibility-level upper limits on the cross-section times acceptance~\cite{EXOT-2010-07} for the signals described above. The method uses a constant prior for the signal cross-section and Gaussian priors for nuisance parameters corresponding to systematic uncertainties. For both observed and expected limits, the sliding window fit is performed for each value of the mass parameter, adding the tested signal shape with a floating normalization to the functional forms stated above. The expected limits are calculated using pseudo-experiments generated from the fit parameters of those functional forms. The uncertainties on the \Zprime{} signal model include the jet energy scale and the luminosity. The impact of the jet energy resolution uncertainty is negligible. For the Gaussian model, a constant jet energy scale uncertainty of 3\% is applied in accordance with the measured impact of this uncertainty on the \Zprime{} samples. The uncertainty in the integrated luminosity is $\pm 2.2$\%, derived following a methodology similar to that detailed in Ref.~\cite{DAPR-2013-01}. The systematic uncertainties in the background estimate include the choice of the fit function and the uncertainty in the fit parameter values, as described above.

\begin{figure}[h!]
\centering
\includegraphics[width=0.5\columnwidth]{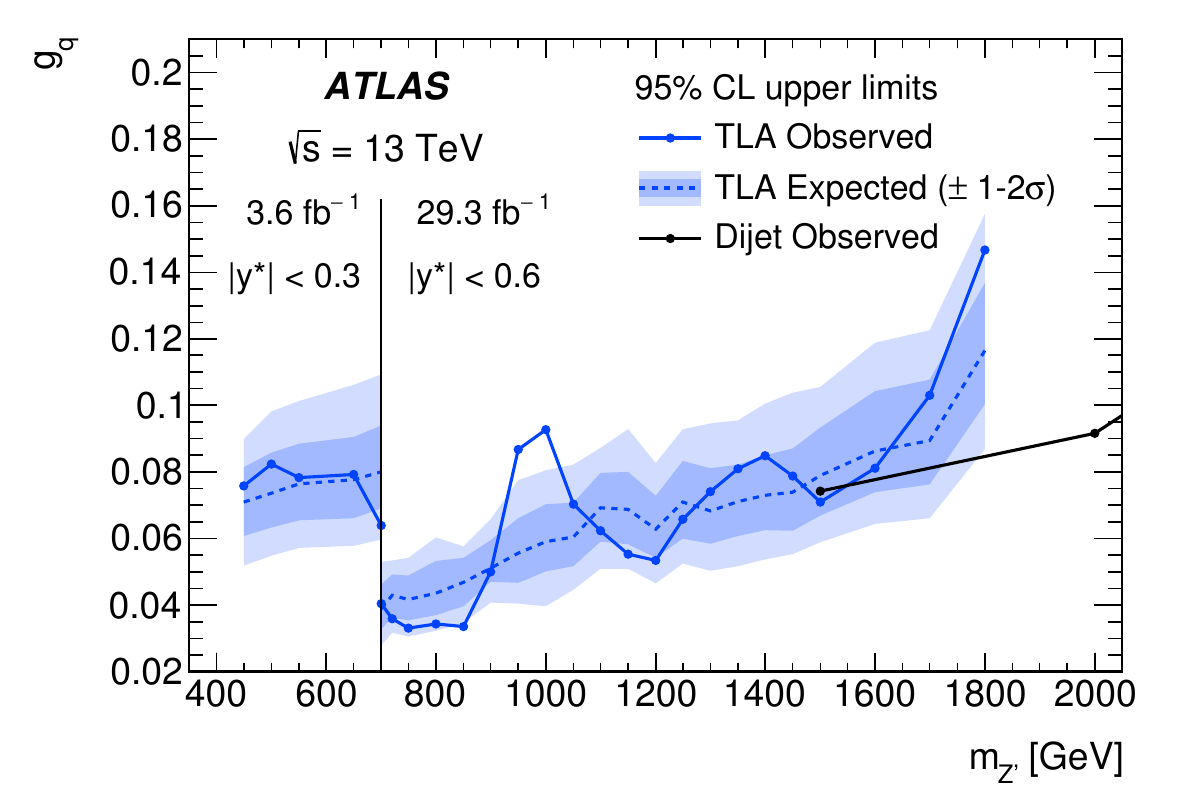}
\caption{The 95\% credibility-level observed and expected upper limits on $g_q$ as a function of $m_{\Zprime}$ for the \Zprime\ model described in the text. The lower-mass part of the limits from Ref.~\cite{EXOT-2016-21} is also shown. Couplings above the solid lines are excluded. The solid and dashed lines represent the observed and expected limits, respectively, and are obtained accounting for the scaling of the signal cross-section with $g_q^2$. The different \ystar\ selections are described in the text.}
\label{fig:zprimeTwoD}
\end{figure}

\begin{figure}
\centering
\includegraphics[width=0.5\columnwidth]{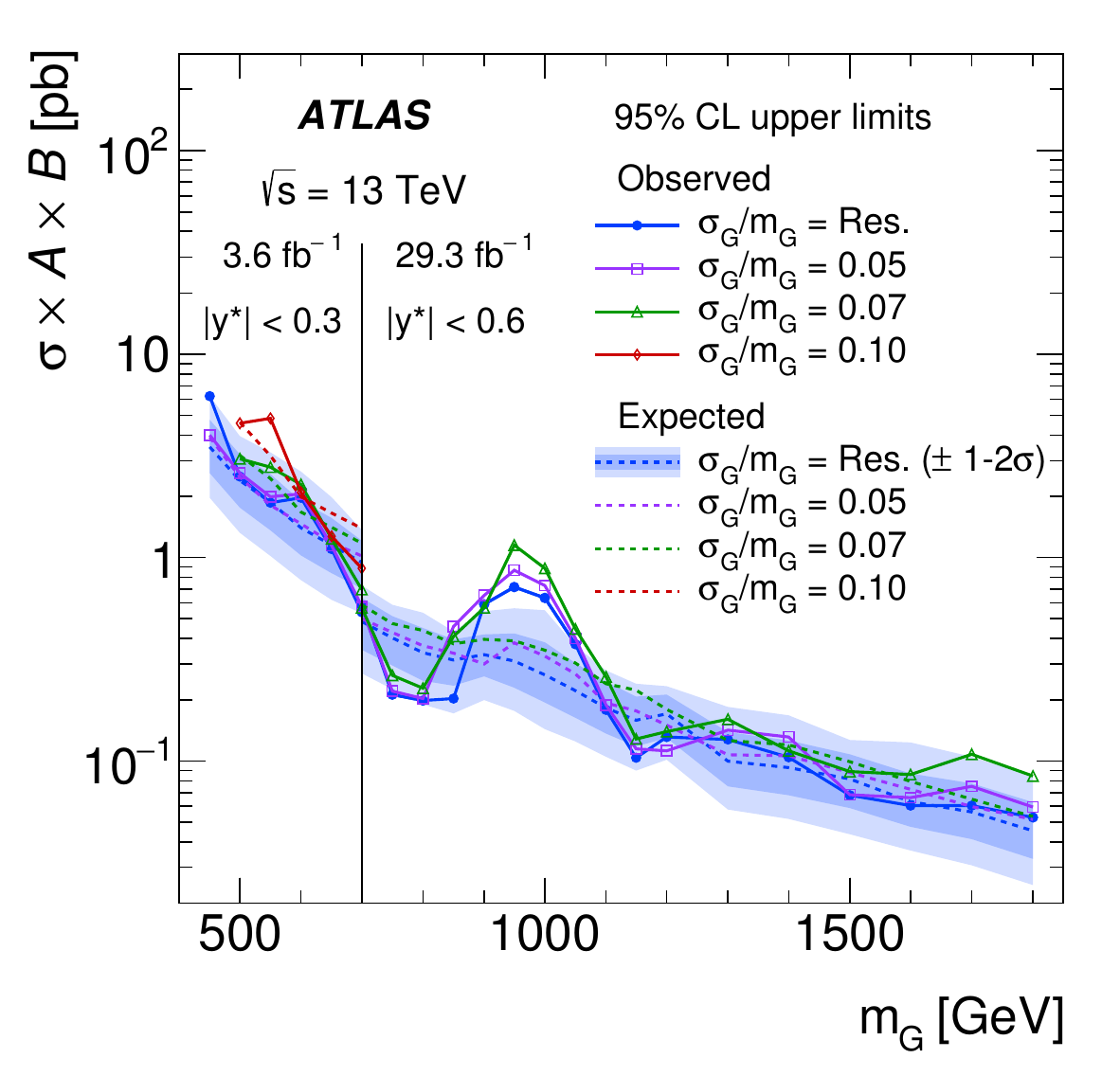}
\caption{The 95\% credibility-level observed upper limits on $\sigma\times A\times B$ for two jets for a hypothetical signal producing a Gaussian contribution to the observed \mjj\xspace distribution. The limits are shown for a relative width $\sigma_G/m_G$ corresponding to a width equal to the detector mass resolution. While the vertical axis is shared by the two selections, the signal acceptance varies, thus the two sets of limit points relate to two different interpretations of $\sigma \times A \times B$ (see text for some typical acceptance values used for models considered by this search). The different \ystar\ selections are described in the text.}
\label{fig:gaussianlimits}
\end{figure}

Figure~\ref{fig:zprimeTwoD} shows limits on the coupling to quarks, $g_q$, as a function of the mass $m_{\Zprime}$ for the \Zprime{} model. Figure~\ref{fig:gaussianlimits} shows limits on a possible Gaussian contribution with a width equal to the detector resolution as a function of the mean mass, $m_G$. In both the \Zprime{} and Gaussian models, upper limits for masses from 450~\GeV\ to 700~\GeV\ are derived using the distribution with $|\yStar| < 0.3$, which is sensitive to the lower masses. Limits for masses above 700~\GeV\ are derived from the \mjj\ distribution with $|\yStar| < 0.6$, except for Gaussian signals with a width of 10\% where only the $|\yStar| < 0.3$ distribution is used.

The limit results show an upward fluctuation at masses of approximately 1~\TeV\ in the $|\yStar| < 0.6$ signal region. This is not seen in Figure~\ref{fig:FullUnblinded}; when searching for excesses in the data, a background-only hypothesis is used for the sliding window fit. In the observed and expected limits, the fit includes the signal shape in addition to the background parameterization as explained above, and can adapt to local data fluctuations that mimic a signal shape. The $|\yStar| < 0.6$ signal region, which uses a smaller sliding-window size, is especially sensitive to the difference in the two approaches. Therefore, limits were not set on signals with a width of 10\% for the $|\yStar| < 0.6$ signal region as the signal is too wide for the sliding-window size.

\FloatBarrier

\section{Conclusions}

In conclusion, this analysis searches for resonances with masses between 450~\GeV{} and 1800~\GeV{} in dijet events using \triggerlevel jets in \integLumi{} of $\sqrt{s}=13$~\TeV\ proton--proton collision data recorded by the ATLAS detector at the LHC. The invariant mass distribution presents no significant local excesses compared to the estimated SM background. This analysis provides 95\% credibility-level limits on \Zprime{} signals and cross-sections for new processes that would produce a Gaussian contribution to the dijet mass distribution. Over much of the mass range, the sensitivity to the coupling to quarks, $g_q$, is improved by a factor of two or more compared to pre-LHC and $\sqrt{s}=8$ and 13~\TeV\ ATLAS results, and is comparable to CMS searches at $\sqrt{s}=8$ and 13~\TeV\ using this technique.
Gaussian contributions with effective cross-sections times acceptance ranging from approximately 6.5~pb at 450~\GeV, to 0.4~pb at 700~\GeV, to 0.05~pb at 1800~\GeV{} are excluded.

%
%
%

%

%
%

%
%
We thank CERN for the very successful operation of the LHC, as well as the
support staff from our institutions without whom ATLAS could not be
operated efficiently.

We acknowledge the support of ANPCyT, Argentina; YerPhI, Armenia; ARC, Australia; BMWFW and FWF, Austria; ANAS, Azerbaijan; SSTC, Belarus; CNPq and FAPESP, Brazil; NSERC, NRC and CFI, Canada; CERN; CONICYT, Chile; CAS, MOST and NSFC, China; COLCIENCIAS, Colombia; MSMT CR, MPO CR and VSC CR, Czech Republic; DNRF and DNSRC, Denmark; IN2P3-CNRS, CEA-DRF/IRFU, France; SRNSFG, Georgia; BMBF, HGF, and MPG, Germany; GSRT, Greece; RGC, Hong Kong SAR, China; ISF, I-CORE and Benoziyo Center, Israel; INFN, Italy; MEXT and JSPS, Japan; CNRST, Morocco; NWO, Netherlands; RCN, Norway; MNiSW and NCN, Poland; FCT, Portugal; MNE/IFA, Romania; MES of Russia and NRC KI, Russian Federation; JINR; MESTD, Serbia; MSSR, Slovakia; ARRS and MIZ\v{S}, Slovenia; DST/NRF, South Africa; MINECO, Spain; SRC and Wallenberg Foundation, Sweden; SERI, SNSF and Cantons of Bern and Geneva, Switzerland; MOST, Taiwan; TAEK, Turkey; STFC, United Kingdom; DOE and NSF, United States of America. In addition, individual groups and members have received support from BCKDF, the Canada Council, CANARIE, CRC, Compute Canada, FQRNT, and the Ontario Innovation Trust, Canada; EPLANET, ERC, ERDF, FP7, Horizon 2020 and Marie Sk{\l}odowska-Curie Actions, European Union; Investissements d'Avenir Labex and Idex, ANR, R{\'e}gion Auvergne and Fondation Partager le Savoir, France; DFG and AvH Foundation, Germany; Herakleitos, Thales and Aristeia programmes co-financed by EU-ESF and the Greek NSRF; BSF, GIF and Minerva, Israel; BRF, Norway; CERCA Programme Generalitat de Catalunya, Generalitat Valenciana, Spain; the Royal Society and Leverhulme Trust, United Kingdom.

The crucial computing support from all WLCG partners is acknowledged gratefully, in particular from CERN, the ATLAS Tier-1 facilities at TRIUMF (Canada), NDGF (Denmark, Norway, Sweden), CC-IN2P3 (France), KIT/GridKA (Germany), INFN-CNAF (Italy), NL-T1 (Netherlands), PIC (Spain), ASGC (Taiwan), RAL (UK) and BNL (USA), the Tier-2 facilities worldwide and large non-WLCG resource providers. Major contributors of computing resources are listed in Ref.~\cite{ATL-GEN-PUB-2016-002}.
%


%

%
\printbibliography

\clearpage
 
\begin{flushleft}
{\Large The ATLAS Collaboration}

\bigskip

M.~Aaboud$^\textrm{\scriptsize 34d}$,    
G.~Aad$^\textrm{\scriptsize 99}$,    
B.~Abbott$^\textrm{\scriptsize 124}$,    
O.~Abdinov$^\textrm{\scriptsize 13,*}$,    
B.~Abeloos$^\textrm{\scriptsize 128}$,    
S.H.~Abidi$^\textrm{\scriptsize 165}$,    
O.S.~AbouZeid$^\textrm{\scriptsize 143}$,    
N.L.~Abraham$^\textrm{\scriptsize 153}$,    
H.~Abramowicz$^\textrm{\scriptsize 159}$,    
H.~Abreu$^\textrm{\scriptsize 158}$
% ...
\bigskip
\\

% ...
$^{13}$Institute of Physics, Azerbaijan Academy of Sciences, Baku; Azerbaijan.\\
% ...
$^{34}$$^{(a)}$Facult\'e des Sciences Ain Chock, R\'eseau Universitaire de Physique des Hautes Energies - Universit\'e Hassan II, Casablanca;$^{(b)}$Centre National de l'Energie des Sciences Techniques Nucleaires (CNESTEN), Rabat;$^{(c)}$Facult\'e des Sciences Semlalia, Universit\'e Cadi Ayyad, LPHEA-Marrakech;$^{(d)}$Facult\'e des Sciences, Universit\'e Mohamed Premier and LPTPM, Oujda;$^{(e)}$Facult\'e des sciences, Universit\'e Mohammed V, Rabat; Morocco.\\
% ...
$^{99}$CPPM, Aix-Marseille Universit\'e, CNRS/IN2P3, Marseille; France.\\
% ...
$^{124}$Homer L. Dodge Department of Physics and Astronomy, University of Oklahoma, Norman OK; United States of America.\\
% ...
$^{128}$LAL, Universit\'e Paris-Sud, CNRS/IN2P3, Universit\'e Paris-Saclay, Orsay; France.\\
% ...
$^{143}$Santa Cruz Institute for Particle Physics, University of California Santa Cruz, Santa Cruz CA; United States of America.\\
% ...
$^{153}$Department of Physics and Astronomy, University of Sussex, Brighton; United Kingdom.\\
% ...
$^{158}$Department of Physics, Technion, Israel Institute of Technology, Haifa; Israel.\\
$^{159}$Raymond and Beverly Sackler School of Physics and Astronomy, Tel Aviv University, Tel Aviv; Israel.\\
% ...
$^{165}$Department of Physics, University of Toronto, Toronto ON; Canada.\\
% ...
$^{a}$ Also at  Department of Physics, University of Malaya, Kuala Lumpur; Malaysia.\\
$^{b}$ Also at Borough of Manhattan Community College, City University of New York, NY; United States of America.\\
$^{c}$ Also at Centre for High Performance Computing, CSIR Campus, Rosebank, Cape Town; South Africa.\\
$^{d}$ Also at CERN, Geneva; Switzerland.\\
$^{e}$ Also at CPPM, Aix-Marseille Universit\'e, CNRS/IN2P3, Marseille; France.\\
% ...
$^{*}$ Deceased

\end{flushleft}


\end{document}